\documentstyle[12pt]{article}                    %%%
\newcommand{\beq}{\begin{equation}}             %%
\newcommand{\eeq}{\end{equation}}               %%
\newcommand{\bqry}{\begin{eqnarray}}            %%
\newcommand{\eqry}{\end{eqnarray}}              %%
\newcommand{\bqryn}{\begin{eqnarray*}}          %%
\newcommand{\eqryn}{\end{eqnarray*}}            %%
\newcommand{\NL}{\nonumber \\}                  %%
\newcommand{\preprint}[1]{\begin{table}[t]      %%
            \begin{flushright}                  %%
            \begin{large}{#1}\end{large}        %%
            \end{flushright}                    %%
            \end{table}}                        %%
\newcommand{\PD}[2]                             %%
    {\frac{\partial^{#2}}{\partial #1^{#2}}}    %%
\begin{document}
\preprint{LA-UR-99-1680}
\title{String Model for Analytic Nonlinear \\ Regge Trajectories}
\author{\\ L. Burakovsky\thanks{E-mail: BURAKOV@T5.LANL.GOV} \
 \\  \\  Theoretical Division, MS B283 \\  Los Alamos National Laboratory \\ 
Los Alamos, NM 87545, USA }
\date{ }
\maketitle
\begin{abstract}
We present a new generalized string model for Regge trajectories $J=J(E^2),$ 
where $J$ and $E$ are the orbital momentum and energy of the string, 
respectively. We demonstrate that this model is not to produce linear 
Regge trajectories, in contrast to the standard Nambu-Goto string, but 
generally nonlinear trajectories, which in many cases can be given in 
analytic form. As an example, we show how the model generates square-root, 
logarithmic and hyperbolic trajectories that have been discussed in the 
literature.
\end{abstract}

\bigskip

\section{Introduction}
In QCD, the characteristic feature of the gluon mediators of the color force 
is their strong self-interaction, because the gluons themselves carry color 
charges. In analogy with the electric lines of force between two electric 
charges, one usually assumes that color charges (quarks) are held together by
color lines of force, but the gluon-gluon interaction pulls these together into
the form of a tube (string). Therefore, the use of the string picture for the
phenomenological description of strong interactions seems to be completely
justified. In fact, the formation of chromoelectric flux tubes between static
quarks is by now a well established feature of lattice QCD simulations 
\cite{lat}. 

The action of the standard relativistic Nambu-Goto string with massive ends 
(quarks) in the parametrization $\tau =t=x^0$ is written as $(\gamma $ is the 
string tension) \cite{BN}
\beq
S=-\gamma \int _{t_1}^{t_2}dt\int _0^\pi d\sigma \sqrt{x'^2(1-\dot{x}^2)+
(\dot{x}x')^2}-\sum_{i=1,2}m_i\int _{t_1}^{t_2}dt\;\!\sqrt{1-\dot{x}_i^2},
\eeq
$$x\equiv {\bf x}={\bf x}(t,\sigma ),\;\;\;x_i\equiv x(t,\sigma _i),
\;\;i=1,2,\;\;\sigma _1=0,\;\;\sigma _2=\pi ,$$
where from now on the dot and the prime stand for the derivative with respect 
to $t$ and $\sigma ,$ respectively, unless otherwise specified. The action (1)
is invariant under arbitrary transformations of $\sigma ,$ therefore the 
string Lagrangian in Eq. (1) is degenerate, The corresponding constraint among
the canonical variables $x(t,\sigma )$ and $p(t,\sigma )$ $(p(t,\sigma )\equiv
{\bf p}(t,\sigma ))$ is
\beq
px'=0,
\eeq
for
\beq
p=p(t,\sigma )=\frac{\partial L_{str}}{\partial \dot{x}}=\frac{\gamma ^2}{
L_{str}}\;\!\Big[ x'(\dot{x}x')^2-\dot{x}(x')^2\Big] ,
\eeq
$$L_{str}=-\gamma \sqrt{x'^2(1-\dot{x}^2)+(\dot{x}x')^2}.$$
With the equalities
\beq
p\dot{x}-L_{str}=-\frac{\gamma ^2}{L_{str}}\;\!x'^2,\;\;\;p^2+\gamma ^2x'^2=
\left( -\frac{\gamma ^2}{L_{str}}\;\!x'^2\right) ^2,
\eeq
one obtains the expression for the canonical Hamiltonian, as follows,
\bqry
H & = & \int _0^\pi d\sigma \sqrt{p^2+\gamma ^2x'^2}+\sum _{i=1,2}\sqrt{p_i^2+
m_i^2} \NL 
 & = & \int _0^\pi d\sigma |p+\gamma x'|+\sum _{i=1,2}\sqrt{p_i^2+m_i^2},
\eqry
where $p_i=m\dot{x}_i/\sqrt{1-\dot{x}_i^2}.$ 

In the nonrelativistic limit, $|\dot{x}(t,\sigma )|\ll 1,$ $|\dot{x}_i|\ll 1,$
Eq. (1) transforms into \cite{BN}
\beq
S=-\gamma \int _{t_1}^{t_2}dt\int _0^\pi d\sigma \sqrt{x'^2}-\sum_{i=1,2}m_i
\int _{t_1}^{t_2}dt+\sum _{i=1,2}\frac{m_i}{2}\int _{t_1}^{t_2}dt\;\!\dot{
x}_i^2.
\eeq
Integral over $\sigma $ gives the length of the string (with the assumption 
that there are no singularities on the string). The variation of the first 
term in Eq. (6) with respect of the string coordinates leads to the 
requirement on the string to have the form of a linear rod connecting the 
massive ends. The effective action that leads to the equations of motion of 
the massive ends is therefore
\beq
S_{eff}=\int _{t_1}^{t_2}dt\left( -\gamma |x_1(t)-x_2(t)|+
\sum _{i=1,2}\frac{m_i\dot{x}_i^2}{2}\right) .
\eeq 
Hence, in the nonrelativistic limit, the string generates a linear potential
between its massive ends: $V(|x_1-x_2|)=\gamma |x_1-x_2|.$ The same result
holds for the string with massive ends in two-dimensional space-time \cite{BN}
where $x(t,\sigma )$ has only one component; therefore $p(t,\sigma )=0,$ via
(3), and, in view of (5), 
$$H=\gamma \int _0^\pi d\sigma |x'|+\sum _{i=1,2}\sqrt{p_i^2+m_i^2}=
\gamma |x(0)-x(\pi )|+\sum _{i=1,2}\sqrt{p_i^2+m_i^2}.$$

The string model with constant tension is known to predict linearly rising
Regge trajectories $J=J(E^2)$ $(J$ and $E$ are the orbital momentum and energy
of the string, respectively): $J=E^2/(2\pi \gamma )$ \cite{BN}. The string
trajectories are exactly linear in the case of the massless ends, and 
asymptotically linear in the case of the massive ends having some curvature
in the region $E\stackrel{>}{\sim }m_1+m_2.$ The same picture of linear 
trajectories arises from a linear confining potential \cite{KS}.
However, the realistic Regge trajectories extracted from data are {\it 
nonlinear.} Indeed, the straight line which crosses the $\rho $ and $\rho _3$ 
squared masses  corresponds to an intercept $\alpha _\rho (0)=0.48,$ whereas 
the physical intercept is located at 0.55, as extracted by Donnachie and 
Landshoff from the analysis of $pp$ and $\bar{p}p$ scattering data in a simple
pole exchange model \cite{DL}. The nucleon Regge trajectory as extracted from
the $\pi ^{+}p$ backward scattering data is \cite{Lyu}
$$\alpha _N(t)=-0.4+0.9\;\!t+\frac{1}{2}\;\!0.25\;\!t^2,$$
and contains positive curvature. Recent UA8 analysis of the inclusive 
differential cross sections for the single-diffractive reactions $p\bar{p}
\rightarrow pX,$ $p\bar{p}\rightarrow X\bar{p}$ at $\sqrt{s}=630$ GeV reveals 
a similar curvature of the Pomeron trajectory \cite{UA8}:
$$\alpha _{I\!\!P}(t)=1.10+0.25\;\!t+\frac{1}{2}\;\!(0.16\pm 0.02)\;\!t^2.$$
An essentially nonlinear $a_2$ trajectory was extracted in ref. \cite{Bol2} 
for the process $\pi ^{-}p\rightarrow \eta n.$ Note that the nonlinearity
of Regge trajectories was also proven on quite general grounds \cite{nonlin}.

Once the nonlinearity of Regge trajectories is an established fact, a number 
of important issues immediately suggest themselves; e.g.,

(i) the understanding of the reasons for the nonlinearity of trajectories,
 
(ii) the development of the model(s) leading to analytic nonlinear Regge 
trajectories. 

Possible reasons for the nonlinearity of trajectories may be related to the
string breaking in QCD. In the absence of dynamical fermions (e.g., in zero
temperature quenched QCD simulations) does not allow the screening of the
potential between the heavy (static) $Q\bar{Q}$ by virtual color-singlet
light $q\bar{q}$ pairs, and thus the interquark potential is expected to grow
linearly with the separation $R$ for arbitrarily large $R.$ With dynamical
fermions, the static $Q\bar{Q}$ meson can decay into two heavy-light mesons.
Ignoring meson-meson interactions, one can expect the QCD string to break as
soon as the potential exceeds twice the heavy-light mass, i.e., at about 1.5 
fm. (Neglecting quark mass effects on the dynamics of the binding problem, 
which is a reasonable assumption once this mass is small compared to a typical 
binding energy of $\sim 500$ MeV, the string breaking distance should be 
shifted by $\triangle r\approx 2\triangle m/\sigma _{eff}$ when changing the 
quark mass by $\triangle m,$ where $\sigma _{eff}$ is the (effective) string 
tension, to be discussed explicitly below.) The expectation of the string
breaking due to the color screening described above has indeed been confirmed
by some of the data on QCD lattice simulations \cite{lat2}.

The purpose of the present letter is to present a model for nonlinear Regge 
trajectories. This model is the generalization of the standard string model.
Indeed, if the string breaking really happens in QCD, as should become clear 
with the forthcoming sophisticated lattice measurements, there is a 
necessity to reconsider the canonical string model and modify the notion of 
a constant string tension, in order to include the color screening effects.
Before we proceed with the presentation of such a modified string model, let
us note that potential models, in general, lead to nonlinear trajectories 
(for potentials that are different from a linearly rising one), but these 
trajectories cannot generally be cast into analytic form \cite{traj}. The 
model we shall present in what follows in many cases leads to {\it analytic} 
nonlinear Regge trajectories. 

\section{Generalized string model}
Here we wish to generalize the standard string formulation reviewed above on
the case of arbitrary potential between the string massive ends. Such 
generalization is done by the modification of the standard string tension into
the effective string tension which is a function of $|x|,$ as follows:
\beq
S_{gen}=-\int _{t_1}^{t_2}dt\int _0^\pi d\sigma \;\!\gamma (|x|)\sqrt{
x'^2(1-\dot{x}^2)+(\dot{x}x')^2}-\sum_{i=1,2}m_i\int _{t_1}^{t_2}dt\;\!
\sqrt{1-\dot{x}_i^2}.
\eeq
As we explain below, in the nonrelativistic limit the effective string tension
is the derivative of the interaction potential between the massive ends of the
string. Therefore, different choices of the effective string tension would be
related to different potentials, which makes it possible to deal, among the 
others, with color-screened potentials, i.e., potentials that approach 
constant values at large separations; e.g., $V(\rho )=\gamma /\mu \;\!(1-\exp 
(-\mu \rho ))$ which is used to fit the lattice QCD data (e.g., first paper 
of \cite{lat2}), and two examples in Section 3. As is clear from the above
discussion, such color-screened potentials may be relevant to QCD.

Obviously, the new action (8) is again invariant under arbitrary 
transformations of $\sigma ,$ and therefore the constraint (2), and Eqs. 
(3)-(5) hold in this case as well, with $\gamma =\gamma (|x|).$ In the 
nonrelativistic limit, however, in place of (7) one will now obtain
$$S_{gen,eff}=\int _{t_1}^{t_2}dt\left( -\int _0^\pi d\sigma \;\!
\gamma (|x|)|x'|+\sum _{i=1,2}\frac{m_i\dot{x}_i^2}{2}\right) $$
\beq
=\int _{t_1}^{t_2}dt\left( -V(|x_1-x_2|)+\sum _{i=1,2}\frac{m_i\dot{
x}_i^2}{2}\right) .
\eeq
In contrast to the previous case of the standard string, it is seen in the
above relations that now
\beq
\gamma (|x|)=\frac{1}{|x'|}\;\!\frac{dV(|x|)}{d\sigma }=\frac{dV(|x|)}{d|x|},
\eeq
i.e., the effective string tension is the derivative of the potential with
respect to the distance. Obviously, in the case of a linear potential, the
effective string tension reduces to the standard (constant) one. 

It should be noted that the other generalizations of the standard string have 
also been discussed in the literature \cite{gen,wiggly}. Specifically, refs. 
\cite{gen} deal with the linearized string Lagrangian that generally has 
different coefficients of $\dot{x}^2$ and $x'^2,$ which are to be associated 
with the string mass density and tension, respectively. Both of these 
coefficients are functions of $\sigma .$ Refs. \cite{wiggly} deal with the 
so-called wiggly string (membrane) for which the string mass density and 
tension are generally different also. With $\gamma $ being a function of 
$\sigma $ not $x,$ as in \cite{gen}, (i) the invariance of the action (8) 
under arbitrary transformations of $\sigma $ disappears, and (ii) the 
interpretation of $\gamma $ as an effective string tension, via Eq. (10), 
is lost, as seen in (9). Also, there are claims in the literature that the
choice of the string mass density and tension being different from each other 
contradicts the principles of relativity which require that the both coincide 
\cite{BI}. In our case (upon the linearization of the Lagrangian in Eq. (8))
both the string mass density and tension would be equal to $\gamma (|x|).$

Similarly to the standard case of a constant string tension which represents 
the relativization {\it \`{a} la} Poincar\'{e} of the nonrelativistic two-body 
problem with linear potential \cite{Shav}, the generalized string that we are
discussing here can be considered as the relativization of a nonrelativistic 
two-body problem with arbitrary potential.   

\subsection{The dynamics of the generalized string model}
By varying the action of the generalized string with massive ends 
(here the dot stands for the derivative with respect to $\tau ),$
\beq
S_{gen}=\int \!\!\!\!\int d\tau d\sigma L(x,\dot{x},x')+\sum _{i=1,2}
L^{(m)}(\dot{x}_i),
\eeq
one obtains the equations of motion of the generalized string,
\beq
\frac{d}{d\tau }\frac{\partial L}{\partial \dot{x}}+\frac{d}{d\sigma }
\frac{\partial L}{\partial x'}=\frac{\partial L}{\partial x},
\eeq
and the boundary conditions which represent the equations of motion of the
massive ends:
\beq
\frac{d}{d\tau }\frac{\partial L^{(m)}}{\partial \dot{x}_i}=\frac{
\partial L}{\partial x'},\;\;\;x=x_i.
\eeq

In the gauge $\tau =t$ discussed above, the equation of motion of the 
generalized string reduce to
\beq
\frac{d}{dt}\frac{\partial L}{\partial \dot{x}}+\frac{d}{d\sigma }\frac{
\partial L}{\partial x'}=\frac{\partial L}{\partial x},\;\;\;x\equiv {\bf x},
\eeq
and the boundary conditions are
\bqry
m_1\;\!\frac{d}{dt}\frac{\dot{x}_1}{\sqrt{1-\dot{x}_1^2}} & = & \frac{
\partial L}{\partial x'},\;\;\;\sigma =0, \NL
m_2\;\!\frac{d}{dt}\frac{\dot{x}_2}{\sqrt{1-\dot{x}_2^2}} & = & \frac{
\partial L}{\partial x'},\;\;\;\sigma =\pi . 
\eqry

Let us show that, similarly to the standard case of the string with constant
tension, there are solutions to the equations of motion of the generalized
string in the form of a rigid rod connecting the massive ends and rotating 
with frequency $\omega $ about its center of mass:
\beq
x(t,\sigma )=\rho (\sigma )\Big( \cos (\omega t),\;\sin (\omega t),\;0\Big) .
\eeq
Indeed, $\gamma =\gamma (|x|)=\gamma (\rho ),$ since $x^2=\rho ^2;$ therefore 
$d\rho/dx=x/\rho =(\cos (\omega t),\;\sin (\omega t),\;0),$ and $d\gamma /dx=d
\gamma /d\rho $ $d\rho /dx=d\gamma /d\rho $ $(\cos (\omega t),\;\sin (\omega 
t),\;0).$ Hence 
\beq
\frac{\partial L}{\partial x}=\frac{\partial L}{\partial \gamma }\;\!
\frac{d\gamma}{dx}=-\frac{d\gamma }{d\rho }\;\!\rho' \sqrt{1-\omega ^2
\rho ^2}\Big( \cos (\omega t),\;\sin (\omega t),\;0\Big) .
\eeq
Since also
\beq
\frac{d}{dt}\;\!\frac{\partial L}{\partial \dot{x}}=-\frac{\gamma \omega ^2
\rho \rho'}{\sqrt{1-\omega ^2\rho ^2}}\;\!\Big( \cos (\omega t),\;\sin 
(\omega t),\;0\Big) ,
\eeq
\beq
\frac{d}{d\sigma }\;\!\frac{\partial L}{\partial x'}=\left( -\frac{d\gamma }{
d\rho }\;\!\rho' \sqrt{1-\omega ^2\rho ^2}+\frac{\gamma \omega ^2\rho \rho'}{
\sqrt{1-\omega ^2\rho ^2}}\right) \Big( \cos (\omega t),\;\sin (\omega t),
\;0\Big) ,
\eeq
(the last relation is obtained via $d\gamma /d\sigma =d\gamma /d\rho $ 
$\rho'),$ one sees that the equations of motion (14) are satisfied. 

One can easily check that for the rotation (16), the energy of the generalized 
string is given, in view of (5), by
\beq
H=\int d\sigma \sqrt{p^2+\gamma ^2x'^2}=\int d\sigma \;\!\frac{\gamma 
\rho'}{\sqrt{1-\omega ^2\rho ^2}}=\int \frac{d\rho \;\!\gamma (\rho )}{
\sqrt{1-\omega ^2\rho ^2}}.
\eeq
Similarly, the orbital momentum of the generalized string is 
\beq
J=J_z=\int d\sigma \left( xp_y-yp_x\right) =\int d\sigma \;\!\frac{\gamma 
\omega \rho ^2\rho'}{\sqrt{1-\omega ^2\rho ^2}}=\int \frac{d\rho \;\!
\gamma (\rho )\omega \rho ^2}{\sqrt{1-\omega ^2\rho ^2}}.
\eeq
Interestingly enough, in his book \cite{Per} Perkins also presents the above 
relations for the energy and orbital momentum of the generalized string. He
does not however derive these relations from the first principles Lagrangian,
as in Eq. (8). 

By adding the contribution of the massive ends, one finally has the 
expressions for the total energy and orbital momentum of the generalized 
string with massive ends:
\beq
E=\int _0^{r_1}\frac{d\rho \;\!\gamma (\rho )}{\sqrt{1-\omega ^2\rho ^2}}+
\int _0^{r_2}\frac{d\rho \;\!\gamma (\rho )}{\sqrt{1-\omega ^2\rho ^2}}+
\frac{m_1}{\sqrt{1-\omega ^2r_1^2}}+\frac{m_2}{\sqrt{1-\omega ^2r_2^2}},
\eeq
\beq
J=\int _0^{r_1}\frac{d\rho \;\!\gamma (\rho )\omega \rho ^2}{\sqrt{1-\omega ^2
\rho ^2}}+\int _0^{r_2}\frac{d\rho \;\!\gamma (\rho )\omega \rho ^2}{\sqrt{1-
\omega ^2\rho ^2}}+\frac{m_1\omega r_1^2}{\sqrt{1-\omega ^2r_1^2}}+\frac{
m_2\omega r_2^2}{\sqrt{1-\omega ^2r_2^2}}.
\eeq
Note that the boundary conditions (15) define the separations of the massive 
ends from the center of mass through the following nonlinear equations:
\beq
\frac{m_i\omega ^2r_i}{\sqrt{1-\omega ^2r_i^2}}=\gamma (r_i)\sqrt{1-\omega ^2
r_i^2},\;\;\;i=1,2.
\eeq

\subsection{Generalized massless string}
The energy and orbital momentum of the generalized massless string, 
$m_1=m_2=0,$ are given by
\beq
E=2\int _0^R\frac{d\rho \;\!\gamma (\rho )}{\sqrt{1-\omega ^2\rho ^2}},\;\;\;
J=2\int _0^R\frac{d\rho \;\!\gamma (\rho )\omega \rho^2}{\sqrt{1-\omega ^2\rho
^2}}, 
\eeq
where $R=1/\omega $ is half of the string length for a given $\omega .$ The
condition $\omega R=1$ follows from, e.g., Eqs. (24) with $m_i\rightarrow 0.$ 

By eliminating $\omega $ from Eqs. (25) one can obtain $J$ as a function of 
$E^2,$ the Regge trajectory. As we demonstrate below, the trajectory is 
generally nonlinear, and in many cases it is given in an analytic form. It 
will be shown elsewhere \cite{prep} that it is possible to uniquely recover 
the potential $(V(r)\sim \int d\rho $ $\gamma (\rho ))$ from the known 
analytic form of Regge trajectory, for both generalized massless and massive 
strings, and the techniques of the corresponding inverse problem will be 
presented in detail. It should be noticed that potential corresponding to an 
analytic Regge trajectory cannot always be recovered itself in an analytic 
form, but only as some power series in $\rho .$ However, in the most important
cases of analytic nonlinear Regge trajectories the corresponding potentials 
happen to be recovered analytically. Below we present such potentials for 
three examples of analytic nonlinear Regge trajectories that have been 
discussed in the literature. For our present purposes, here we present only 
the final results.

\section{Analytic nonlinear Regge trajectories}
In what follows, we consider only the generalized massless string.

\subsection{Square-root trajectory}
The square-root Regge trajectory, $J\propto E_{th}-\sqrt{E_{th}^2-E^2},$ 
where $E_{th}$ is the trajectory energy threshold, has been widely discussed 
by Kobylinsky {\it et al.} \cite{Kob} as the simplest choice of a trajectory 
for dual amplitude with Mandelstam analyticity (DAMA) \cite{DAMA}. It also 
provides for the Orear (small-$t)$ regime of the amplitude. The corresponding 
potential is $(\gamma ,\mu ={\rm const,}$ $V(\rho )\rightarrow \gamma /2\mu $ 
as $\rho \rightarrow \infty ,$ and hence $E\rightarrow E_{th}=\gamma /\mu )$
\beq
V(\rho )=\frac{\gamma }{\pi \mu }\arctan (\pi \mu \rho ),
\eeq
for which
\beq
\gamma (\rho )=\frac{dV(\rho )}{d\rho }=\frac{\gamma }{1+(\pi \mu \rho )^2},
\eeq
and leads, via (25), to
\beq
E=\frac{\pi \gamma }{\sqrt{\omega ^2+\pi ^2\mu ^2}},\;\;\;J=\frac{\gamma }{
\pi \mu ^2}\left( 1-\frac{\omega }{\sqrt{\omega ^2+\pi ^2\mu ^2}}\right) .
\eeq
Eliminating $\omega $ from the above relations gives
\beq
J=\frac{1}{\pi \mu }\left( \gamma /\mu -\sqrt{(\gamma /\mu )^2-E^2}\right) ,
\eeq
i.e., the square-root Regge trajectory. For $E\ll \gamma /\mu ,$ it reduces
to an (approximate) linear trajectory, $J\simeq E^2/(2\pi \gamma ).$ Note 
that in the corresponding quantum case where $J$ takes on integer values only
and the trajectory would develop a (nonzero) intercept (the quantum defect), 
$J(0),$ there would be a finite number of states lying on the trajectory, 
with orbital momenta $0\leq J\leq J_{max}\leq J(0)+\gamma /(\pi \mu ^2).$

\subsection{Logarithmic trajectory}
The logarithmic Regge trajectory, $J\propto -\log (1-E^2/E_{th}^2),$ is the
ingredient of dual amplitude with logarithmic trajectories (DALT) \cite{DALT}.
There are certain theoretical reasons \cite{log} to consider logarithmic 
trajectories that, for large $-t(=-E^2),$ are compatible with fixed angle 
scaling behavior of the amplitude. The corresponding potential is (again 
$\gamma ,\mu ={\rm const,}$ $V(\rho )\rightarrow \gamma /2\mu $ as $\rho 
\rightarrow \infty ,$ and hence $E\rightarrow E_{th}=\gamma /\mu )$
\beq
V(\rho )=\frac{\gamma }{2\pi \mu }\left( 2\arctan (2\pi \mu \rho )-\frac{
\log [1+(2\pi \mu \rho )^2]}{2\pi \mu \rho }\right) ,
\eeq
for which 
\beq
\gamma (\rho )=\gamma \;\!\frac{\log [1+(2\pi \mu \rho )^2]}{
(2\pi \mu \rho )^2},
\eeq 
and leads to
\beq
E=\frac{\gamma }{2\pi \mu ^2}\left( \sqrt{\omega ^2+4\pi ^2\mu ^2}-\omega 
\right),\;\;\;J=\frac{\gamma }{2\pi \mu ^2}\;\!\log \frac{\omega +\sqrt{
\omega ^2+4\pi ^2\mu ^2}}{2\omega },
\eeq
from which eliminating $\omega $ (viz., $\omega =\pi (\gamma ^2-\mu ^2E^2)/(
\gamma E))$ gives
\beq
J=-\frac{\gamma }{2\pi \mu ^2}\;\!\log \left( 1-\frac{E^2}{(\gamma /\mu )^2}
\right) ,
\eeq
i.e., the logarithmic Regge trajectory. For $E\ll \gamma /\mu ,$ it again
reduces to an (approximate) linear form, $J\simeq E^2/(2\pi \gamma ).$ Note 
that, in contrast to the previous example of square-root trajectory, although 
the trajectory has an energy threshold, in the corresponding quantum case the 
number of states lying on the trajectory would be infinite.

In each of the two examples considered above, the potential belongs to the 
family of the color-screened potentials mentioned in Section 2. (The exact 
form of the color-screened potential, if it is realized in QCD, will be
discussed elsewhere.) Note that with a different choice of signs in Eqs. 
(29),(33) (e.g., $J=\gamma /(2\pi \mu ^2) \log [1+E^2/(\gamma /\mu )^2])$ 
there would be a difficulty with the analytic continuation of the trajectory
in the region $E^2(=t)<0,$ since in this region the trajectory cannot have 
imaginary part. Note also that besides the trajectories (29),(33), a 
hybrid square-root--logarithmic trajectory, $J\propto -\log (1+\beta \sqrt{
E_{th}^2-E^2}),$ has been motivated and discussed in \cite{log,Laszlo}. We 
have not menaged to obtain the analytic form of the corresponding potential. 

\subsection{Hyperbolic trajectory}
The hyperbolic trajectory, $J\propto \cosh (E)-1,$ results in the 
$\kappa $-deformed Poincar\'{e} phenomenology \cite{kappa}. Here the 
corresponding potential is
\beq
V(\rho )=\frac{\gamma \rho \;\!\Phi (-(\pi \mu \rho )^2,\;\!2,\;\!1/2)}{4}=
\gamma \rho \sum _{k=0}^\infty \frac{[-(\pi \mu \rho )^2]^k}{(2k+1)^2}=
\gamma \rho \left( 1-\frac{(\pi \mu \rho )^2}{9}+\frac{(\pi \mu \rho )^
4}{25}-\ldots \right) ,
\eeq
where 
\beq
\Phi (z,\;\!s,\;\!a)=\sum _{k=0}^\infty \frac{z^k}{(a+k)^s}
\eeq
is the so-called Lerch's transcendent. The corresponding $\gamma (\rho ),$
\beq
\gamma (\rho )=\gamma \;\!\frac{\arctan (\pi \mu \rho )}{\pi \mu \rho },
\eeq
leads to
\beq
E=\frac{\gamma }{\mu }\;\!\log \left( \pi \mu /\omega +\sqrt{1+(\pi \mu 
/\omega )^2}\right) ,
\eeq
\beq
J=\frac{\gamma }{\pi \mu ^2}\left( \sqrt{1+(\pi \mu /\omega )^2}-1\right) ,
\eeq
from which eliminating $\omega $ $(\pi \mu /\omega =\sinh (\mu E/\gamma ))$ 
gives
\beq
J=\frac{\gamma }{\pi \mu ^2}\Big[ \cosh \left( \frac{E}{\gamma /\mu }
\right) -1\Big] ,
\eeq
i.e., the hyperbolic trajectory. For $E\ll \gamma /\mu ,$ it reduces to the
linear form, as well as in the above two cases: $J\simeq E^2/(2\pi \gamma ).$

\section{Concluding remarks}
We have presented a new generalized string model that leads to nonlinear Regge
trajectories which in many cases can be given in analytic form. We have 
considered three examples of how this model generates square-root, logarithmic
and hyperbolic trajectories that have been discussed in the literature. 

The main phenomenological implication of the model presented here is the 
possibility to obtain Regge trajectory for an arbitrary (nonrelativistic) 
potential in general, and for a color-screened potential in particular. Since 
trajectory for the latter is characterized by an energy threshold, and in some
cases by a finite number of states, as in the example of square-root 
trajectory considered, it may be of relevance to QCD to predict the numbers 
of states lying on different trajectories, and the corresponding energy 
thresholds, if a color-screened potential of, e.g., the type (26) is indeed 
realized in QCD. 

Of certain interest would be further exploration of the generalized string 
model discussed here, and the applications of this model to the hadron 
scattering (the form of the corresponding scattering amplitude) and the 
thermodynamics of hot hadronic matter (the equation of state of the ensemble
of the generalized strings), as well as the quantization of this model. These,
and related issues, e.g., the understanding of the form of the color-screened 
potential,  if is realized in QCD, will be the subjects of further study, to 
be undertaken elsewhere. 

\section*{Acknowledgements}
The author wishes to thank M.M. Brisudov\'{a}, T. Goldman, L.P. Horwitz and
P.R. Page for very valuable discussions during the preparation of this work.

%\bigskip

\end{document}